\documentstyle[aps,prl,preprint,floats,epsfig]{revtex}

\begin{document}

\preprint{\tighten\vbox{\hbox{\hfil CLNS 98/1573}
                        \hbox{\hfil CLEO 98-10}
}}

\title
{\Large\bf Study of 3-prong hadronic $\tau$ decays with charged kaons}

% Your author list ***DOES NOT*** go here!
% is goes below where you are instructed to insert it...
\author{CLEO Collaboration}
\date{\today}

\maketitle
\tighten

\begin{abstract} 

 Using a sample of 4.7 $fb^{-1}$ integrated luminosity accumulated
with the CLEO-II detector at the Cornell Electron Storage Ring (CESR),
we have measured the ratios of branching fractions
${\cal B}(\tau^- \rightarrow K^- h^+ \pi^- \nu_{\tau})/
   {\cal B}(\tau^- \rightarrow h^- h^+ h^- \nu_{\tau})=
(5.16\pm 0.20 \pm 0.50)\times 10^{-2}$,
${\cal B}(\tau^-\rightarrow K^-h^+\pi^-\pi^0\nu_\tau)/
    {\cal B}(\tau^-\rightarrow h^-h^+h^-\pi^0\nu_\tau)=
(2.54\pm 0.44 \pm 0.39)\times 10^{-2}$, 
 ${\cal B}(\tau^-\rightarrow K^-K^+\pi^-\nu_\tau)/
      {\cal B}(\tau^-\rightarrow h^-h^+h^-\nu_\tau)=
(1.52\pm 0.14 \pm 0.29)\times 10^{-2}$, and
the upper limit: 
 ${\cal B}(\tau^-\rightarrow K^-K^+\pi^-\pi^0\nu_\tau)/
    {\cal B}(\tau^-\rightarrow h^-h^+h^-\pi^0\nu_\tau)<0.0154$ at 95\% C.L.
Coupled with additional experimental information, we use our results to
extract information on the structure of three-prong tau decays to charged
kaons.
\end{abstract}
\newpage

{
\renewcommand{\thefootnote}{\fnsymbol{footnote}}
\begin{center}
S.~J.~Richichi,$^{1}$ H.~Severini,$^{1}$ P.~Skubic,$^{1}$
A.~Undrus,$^{1}$
M.~Bishai,$^{2}$ S.~Chen,$^{2}$ J.~Fast,$^{2}$
J.~W.~Hinson,$^{2}$ N.~Menon,$^{2}$ D.~H.~Miller,$^{2}$
% ...
\end{center}
 
\small
\begin{center}
$^{1}${University of Oklahoma, Norman, Oklahoma 73019}\\
$^{2}${Purdue University, West Lafayette, Indiana 47907}\\
% ...
\end{center}

\setcounter{footnote}{0}
}
\newpage

% Insert body of the text here.

%===========================================  Intro

\section{Introduction}

Decays of the   
$\tau$ lepton present a unique opportunity to confirm
and further probe the Standard Model. The large mass of the $\tau$ lepton
makes possible decays into hadrons in
an environment where the initial state is simple 
and well understood\cite{tsai}. This allows comparison with
hadron production at comparable center of mass energies from
processes such as
pion-nucleon, nucleon-nucleon and electron-positron
collisions. 
Strange $\tau$ decays give us information on $SU(3)_f$
symmetry breaking, and direct measurements of the 
Cabibbo angle ($\theta_c$). Alternately, 
the rates for such Cabibbo-suppressed
decays (e.g., $\tau\to K\nu_\tau, \tau\to K^*\nu_\tau$, and 
$\tau\to K_1\nu_\tau$)
can be predicted using corresponding measurements of the
Cabibbo favored modes
($\tau\to \pi\nu_\tau$, $\tau\to \rho\nu_\tau$, and $\tau\to a_1\nu_\tau$).
For the case of vector coupling, 
the theoretically expected ratio of
branching ratios,
${\cal B}(\tau\to K^*\nu_\tau)/{\cal B}(\tau\to\rho\nu_\tau)$\cite{EQNKST} can
be written
 following the approach of the Das-Mathur-Okubo sum rules\cite{DMO}
as
$f(m_{K^*},m_{\rho},m_{\tau})\tan^{2}\theta_c g_{K^*}^2/g_{\rho}^2$, 
where $f$ is a factor that incorporates the phase space available for decays
into $\rho$ and $K^*$, $\theta_c$ is the Cabibbo angle, and the 
factors $g_{K^*}$ and $g_{\rho}$
reflect the coupling strengths of the $K^*$ and $\rho$ to the vector current.
In the limit of exact SU(3) symmetry, there is no differentiation between the
$K^*$ and the $\rho$, so $g_{K^*}=g_{\rho}$.
In the more realistic case of       
broken symmetry, however, the couplings $g$
are directly proportional to mass.
In this approximation, using $\sin\theta_c$=0.221 \cite{PDG96},
the ratio of rates equals 0.068.
There are many 
experimental results 
for $\tau\to K^*\nu_\tau$
(where the charged $K^*$ is observed through the
very clean decay chain: $K^*\to K^0_S\pi$; $K^0_S\to\pi^+\pi^-$) which are
in agreement
with this prediction\cite{PDG96}.

We expect similar relationships to hold for the coupling of the tau to the
axial vector
$K_1$ resonance relative to
the $a_1$ resonance. Unfortunately, $\tau\to K_1\nu_\tau$ is not as
well-studied as $\tau\to K^*\nu_\tau$, in part due to the fact that 
the $K_1$ decay most often leads to 
multi-prong events that include charged kaons.
Whereas the
$K^*$ can be 
unambiguously identified through its decay, $K^*\to K^0_S\pi$,
a measurement of charged kaons
requires good K/$\pi$ particle identification in order
to separate kaons from the
substantially more numerous pions.
Moreover,
theoretical understanding of charged kaon production in tau decay
is hampered by uncertainties in the 
production mechanism;
there is a wide range of predictions
for the mass spectrum (expected to be dominated by $K_1(1270)$ and
$K_1(1400)$),
the $K^*\pi/K\rho$ ratio (in $K\pi\pi$), and the helicity
amplitudes for $\tau$ decays to kaons.

In recent years the large data samples accumulated
at CLEO and LEP have allowed much-improved measurements of 
inclusive decays of tau leptons to charged kaons, complementing
similar measurements of inclusive decays of tau leptons to neutral kaons.
In this analysis we measure the ratio of branching fractions
of $\tau^-\rightarrow K^-h^+\pi^-(\pi^0)\nu_\tau$ and
$\tau^-\rightarrow K^-K^+\pi^-(\pi^0)\nu_\tau$ 
relative
to $\tau^-\rightarrow h^-h^+h^-(\pi^0)\nu_\tau$,
where
$h^\pm$ can be either a charged pion or kaon.\footnote{Charge conjugate modes
are implied throughout the paper.}
 The decay $\tau\to K^-\pi^+\pi^-(\pi^0)\nu_\tau$ proceeding
 through the $K^-K^0(\pi^0)\nu_\tau$ intermediate state has been measured
 in \cite{kshort_paper}; in our analysis, these are considered  background 
 since we are interested in studying tau decays directly into
 3 or 4 mesons that include charged kaons.

%===========================================  Event selection
\section{Data sample and event selection}

   Our data sample contains approximately 4.3 million $\tau$-pairs
produced in $e^+e^-$ collisions, corresponding to an integrated
luminosity of 4.7 $fb^{-1}$. The data were collected with the CLEO-II
detector\cite{detector} 
at the Cornell Electron Storage Ring operating at a center-of-mass 
energy approximately 10.58 GeV. 

The CLEO~II detector
is a general purpose solenoidal magnet spectrometer and
calorimeter. 
The detector was
designed for efficient triggering and reconstruction of
two-photon, tau-pair, and hadronic events.
Measurements of charged particle momenta are made with
three nested coaxial drift chambers consisting of 6, 10, and 51 layers,
respectively.  These chambers fill the volume from $r$=3 cm to $r$=1 m, with
$r$ the radial coordinate relative to the beam (${\hat z}$) axis. 
This system is very efficient ($\epsilon\ge$98\%) 
for detecting tracks that have transverse momenta ($p_T$)
relative to the
beam axis greater than 200 MeV/$c$, and that are contained within the good
fiducial volume of the drift chamber ($|\cos\theta|<$0.94, with $\theta$
defined as the polar angle relative to the beam axis). 
 The charged particle detection efficiency in the fiducial
volume decreases to 
approximately 90\% at $p_T\sim$100 MeV/$c$. 
For $p_T<$100 MeV/$c$, the efficiency
decreases roughly linearly to zero at a threshold of $p_T\approx$30 MeV/$c$.
This system achieves a momentum resolution of $(\delta p/p)^2 =
(0.0015p)^2 + (0.005)^2$ ($p$ is the momentum, measured in GeV/$c$). 
Pulse height measurements in the main drift chamber provide specific
ionization ($dE/dx$) resolution
of 5.5\% for Bhabha events, giving good $K/\pi$ separation for tracks with
momenta up to 700 MeV/$c$ and separation nearly 2$\sigma$ in the relativistic
rise region above 2 GeV/$c$. 
Outside the central tracking chambers are plastic
scintillation counters, which are used as a fast element in the trigger system
and also provide particle identification information from time-of-flight
measurements.  

Beyond the time-of-flight system is the electro-magnetic calorimeter,
consisting of 7800 thallium-doped CsI crystals.  The central ``barrel'' region
of the calorimeter covers about 75\% of the solid angle and has an energy
resolution which is empirically found to follow:
\begin{equation}
~~~~~~~~~~~~~~~~~~
~~~~~~~~~~~
\frac{ \sigma_{\rm E}}{E}(\%) = \frac{0.35}{E^{0.75}} + 1.9 - 0.1E;
                                \label{eq:resolution1}
\end{equation}
$E$ is the shower energy in GeV. This parameterization includes
effects such as noise, and translates to an
energy resolution of about 4\% at 100 MeV and 1.2\% at 5 GeV. Two end-cap
regions of the crystal calorimeter extend solid angle coverage to about 95\%
of $4\pi$, although energy resolution is not as good as that of the
barrel region. 
The tracking system, time of flight counters, and calorimeter
are all contained 
within a superconducting coil operated at 1.5 Tesla. 
Flux return and tracking
chambers used for muon detection are located immediately outside the coil and 
in the two end-cap regions.

    We select $e^+e^-\to\tau^+\tau^-$ 
events having a ``1vs3'' topology
in which one $\tau$ lepton decays into one charged particle (plus 
possible
neutrals),
and the other
$\tau$ lepton decays into 3 charged hadrons (plus possible neutrals).
An event is separated
into two hemispheres based on the measured
event thrust axis\footnote{The thrust axis 
of an event is chosen so that the sum of longitudinal (relative to this
axis) momenta of all charged tracks has a maximum value.}. Loose
cuts on ionization measured in the drift chamber,  
energy deposited in the calorimeter and the maximum penetration
depth into the muon detector system  are applied
to charged tracks in the signal (3-prong) hemisphere to reject leptons.
Backgrounds from $\tau$ and hadronic events with $K_S^0$ are  suppressed
by requirements on the impact parameters of charged tracks.
  To reduce the background from two-photon collisions 
($e^+e^-\to e^+e^-\gamma\gamma$ with $\gamma\gamma\to$hadrons or
$\gamma\gamma\to l^+l^-$),
cuts on visible energy ($E_{vis}$) and
total event transverse momentum ($P_t$) are applied:
2.5 GeV $< E_{vis}< 10$ GeV, and $P_t>$0.3 GeV/$c$. We also require
the invariant mass of the tracks and showers in the 3-prong hemisphere,
calculated under the $\pi^-\pi^+\pi^-$ hypothesis,
to be less than 1.7 GeV/$c$.

% consistent with true $\tau$ decays:

Events are accepted for
which the tag hemisphere (1-prong side) is consistent with
one of the following four decays:  
$\tau^+\rightarrow e^+ \nu_e {\overline \nu_\tau}$, 
$\tau^+\rightarrow \mu^+ \nu_\mu {\overline \nu_\tau}$,  
$\tau^+\rightarrow \pi^+ {\overline  \nu_\tau}$, or
$\tau^+\rightarrow \rho^+ {\overline \nu_\tau}$. 

For the
$\tau^-\rightarrow K^-h^+\pi^-(\pi^0)\nu_\tau$ analysis, we
determine the kaon and pion
yields, using the
two same-sign tracks from the three-prong hemisphere. For the
$\tau^-\rightarrow K^-K^+\pi^-(\pi^0)\nu_\tau$ mode, only the track
having sign
opposite to its parent $\tau$ is considered as a candidate kaon.
Note that 
we implicitly assume that all signal kaons originating from $\tau$
decays in our selected 1vs3 samples come 
from one of the decay modes 
$\tau^-\rightarrow K^-\pi^+\pi^-\nu_\tau$,
$\tau^-\rightarrow K^-K^+\pi^-\nu_\tau$, 
$\tau^-\rightarrow K^-\pi^+\pi^-\pi^0\nu_\tau$, or
$\tau^-\rightarrow K^-K^+\pi^-\pi^0\nu_\tau$. 
The decays $\tau^-\rightarrow \pi^-K^+\pi^-\nu_\tau$ and
 $\tau^-\rightarrow K^-\pi^+K^-\nu_\tau$ 
are extremely small in the 
Standard Model and have not been experimentally observed, and
the decay rate for $\tau^-\rightarrow K^-K^+K^-\nu_\tau$ is expected be 
 $\sim 1\%$
relative to that for
$\tau^-\rightarrow K^-\pi^+\pi^-\nu_\tau$ due to the limited
phase space and the low probability of $(s\bar{s})$ popping.

Candidate events with and without 
$\pi^0$'s are distinguished by the characteristics of showers in the
electro-magnetic calorimeter.
%----old----
%A `photon' candidate is defined as a shower in the barrel (end-cap) 
%region of the electro-magnetic calorimeter  with energy above 
%$40 MeV$  ($60 MeV$) and having an energy deposition pattern consistent 
%with true photons. It must be separated from the closest charged track 
%by at least 20 degrees. $Kh\pi$ and $KK\pi$ candidates are defined as 
%those events having zero or one photons in the 3-prong hemisphere.
%For the $Kh\pi\pi^0$ and $KK\pi\pi^0$ decay modes there must be at 
%least two photons in the signal hemisphere, provided that no more than 
%two photon candidates have a shower energy above 300 MeV. The two most 
%energetic photons in  this hemisphere are then paired to form $\pi^0$ 
%candidates.
%-----new-----
A `photon' candidate is defined as a shower in the barrel 
region of the electromagnetic calorimeter  with energy above 
40 MeV and having an energy deposition pattern consistent 
with true photons. It must be separated from the closest charged track 
by at least 30 cm (20 cm for photons used in 
$\pi^0$ reconstruction).  $\tau^-\to K^-h^+\pi^-\nu_\tau$ and 
$\tau^-\to K^-K^+\pi^-\nu_\tau$ candidates are defined as 
those events having zero photons with energy above 100 MeV
in the 3-prong hemisphere. For the $\tau^-\to K^-h^+\pi^-\pi^0\nu_\tau$ 
and $\tau^-\to K^-K^+\pi^-\pi^0\nu_\tau$ decay 
modes there must be at  least two photons in the signal hemisphere, but  
no more than  two photon candidates having a shower energy 
above 100 MeV. The two most energetic photons in  this hemisphere are 
then paired to form $\pi^0$  candidates.

\section{ Signal extraction procedure}

      To find the number of events with kaons,
we use specific ionization  
information from the central drift chamber. For each track
we calculate the parameter $\delta_K$, defined as
the deviation of the measured energy loss relative
to that expected for true kaons in units of the measured $dE/dx$ resolution.
For true kaons, this variable
is distributed as a unit Gaussian centered at 0. For pion tracks,
$\delta_k$ also has a Gaussian-like shape but with the mean shifted from
zero in a momentum-dependent manner.
In this analysis we concentrate on those 
tracks having momentum $p>$1.5 GeV/$c$. 
Although the $K/\pi$ separation
is better at low momenta, 
we focus on this high momentum region because the separation
varies only slowly through this regime and the systematics of signal
extraction are therefore more tractable. 
These tracks generally have the highest momenta of the 
tracks in the three-prong
hemisphere and 
are well-separated spatially from the lower momentum tracks.
Non-$\tau$ background as well as the pion background relative to the
desired kaon signal in real $\tau$ decays 
is also smaller at high momenta.
%
%  NOTE - Jim asks if qq-background is primary reason for using
%  only high P and should it be reflected in the text, and I say
%  not exactly, the problem that is even worse is that pi/K ratio
%  is very high at P<0.7 even though separation is large, so
%  we are very dependent on how well tails (>2-3 sigma) are 
%  calibrated.
%

  The number of kaon and pion tracks in the three-prong hemisphere
is found statistically by fitting the
$\delta_k$ distribution
for charged tracks 
in the three-prong hemisphere 
to the sum of the pion and kaon $\delta_k$ shapes.
Since the $K-\pi$ separation is
modest, it is critical that the shape of the 
$\delta_k$ distribution for pions is well understood.
$K_s^0\rightarrow \pi^-\pi^+$ decays provide a very clean sample of
true pions from which this distribution can be determined from data.
For the pion and kaon shapes, we use a Johnson 
distribution\cite{john1,john2} with mean shifted from zero, and
a unit Gaussian 
\footnote{The difference in
kaon yields obtained using a Johnson distribution
rather than a unit Gaussian to represent the kaon $\delta_k$ distribution
is typically less than 1\%.}
centered at zero, respectively.

%    -Jim says the this is obvious and should be removed.
%The kaon
%shape used is a standard unit Gaussian 
%centered at zero
%The fitted 
%area under the kaon curve directly yields the 
%number of kaons and the area under the pion
%curve yields the number of pions. 

\begin{figure}
\vspace{6cm}

%     \special{psfile=/cdat/lnska1/disk3/ik/cbx/summer97/fitexample.ps 
     \includegraphics{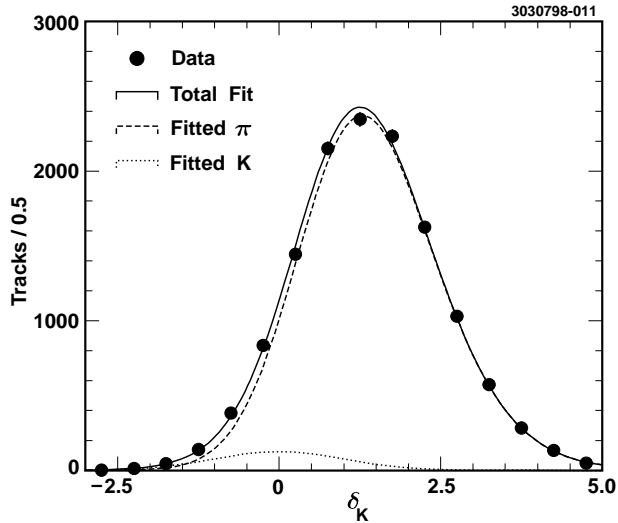}
      \caption{ 
Fit to $\delta_k$ distribution for charged tracks in the 3-prong hemisphere
of candidate $\tau\tau$ events, with the fitted kaon and pion curves overlaid.
Open circles are data, the dotted line shows the kaon contribution to the fit, 
the dashed line shows the  pion contribution and the solid line 
corresponds to the sum of the fitted
kaon and pion curves. The confidence level of the fit is 92\%; 
if the kaon contribution is not included, the C.L. is less than $10^{-3}$.
        \label{exam}}
\end{figure}

Requirements on the minimum number of hits used in the $dE/dx$ calculation
($>20$, out of a maximum of 49) and the polar angle of the candidate kaon track
($|\cos{\theta}|<0.8$, where $\theta$ is the
polar angle of the track relative to the positron beam direction)
ensure that the track is contained in the good fiducial volume of the
drift chamber and that the $dE/dx$ information for the track is of high quality.
Since the $K/\pi$ separation depends on the number of hits ($N_{hit}$), 
as well as momentum ($p$), 
we perform separate $\delta_k$ fits for 36 different bins
in the two parameter ($N_{hit}$, $p$) space.
The kaon and pion yields
for each momentum bin above our minimum
momentum of 1.5 GeV/$c$ are  extracted
knowing the pion (and kaon) $\delta_k$ 
shapes appropriate for each $N_{hit}$ interval over a specified 
momentum range.
An example of a $\delta_k$ fit, showing the kaon and pion components,
is displayed in Fig. \ref{exam}. The $K:\pi$ mixture in this
example is  typical of the 
$\tau^-\to K^-h^+\pi^-\nu_\tau$ analysis 
(of order $1:20$) and is prepared using all tracks with momenta
$1.9$ GeV/$c<p<$2.1 GeV/$c$.

To determine the total number of $\tau\to KX$ events,
we must extrapolate from our measured yields
in the region $p>$1.5 GeV/$c$ to the lower track
momentum region.
This is done by fitting the
measured kaon and pion momentum spectra to the spectra expected
from Monte Carlo (MC) simulations in the $p>$1.5 GeV/$c$ 
region, and integrating over the full  momentum 
range.
Uncertainties in this MC  model  are included in our systematic error.

\begin{figure}

\vspace{8cm}

%    \begin{minipage}[t]{7cm}
         \includegraphics{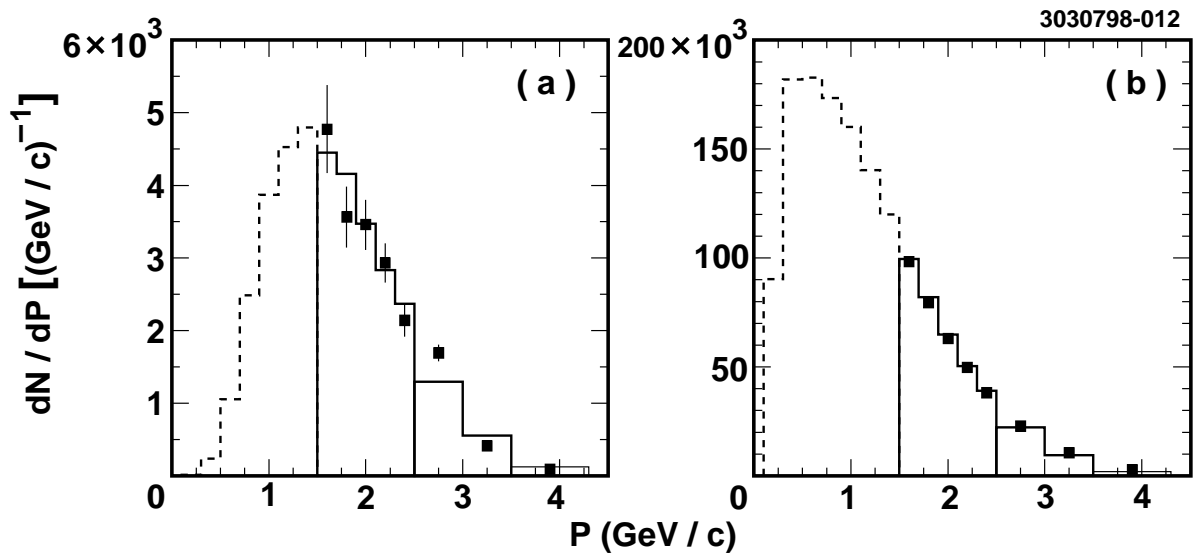}
%   \end{minipage}
%   \hfill
%   \begin{minipage}[t]{7cm}
%        \special{psfile=prd_p.ps 
%        hscale=40 vscale=40 
%        hoffset=-200 voffset=-40 angle=0}
%   \end{minipage}
         \caption{
           Reconstructed a) kaon  and b) pion 
            momentum spectra for  
           $\tau\rightarrow Kh\pi\nu_\tau$ candidates.
         Solid squares are data points and histogram is the MC 
         shape.
         \label{spectrum}}
\end{figure}

The reconstructed momentum spectrum for the
$\tau^-\rightarrow K^-h^+\pi^-\nu_\tau$ 
analysis is shown in 
Fig. \ref{spectrum}, with the  fit
overlaid. For decay modes with $\pi^0$ mesons, $\delta_k$ distributions
are made and fitted separately for cases in which the two-photon invariant
mass falls in the $\pi^0$ signal and sideband regions.
The signal region is taken to be $-4<S_{\gamma\gamma}<3$,
and the sidebands defined as
 $-18<S_{\gamma\gamma}<-10$ and $7<S_{\gamma\gamma}<17$,
where $S_{\gamma\gamma}$ is the number of standard deviations from the
$\pi^0$ mass. 
Subtracting the $K/\pi$
yields from the $\pi^0$ sidebands, the $K/\pi$ signals associated
with true $\pi^0$ production are determined, and the
$\tau^-\rightarrow K^-h^+\pi^-\pi^0\nu_\tau$ yield is extracted.
In Table \ref{tab:numbers} we summarize the total yields and backgrounds
for all four samples.

\begin{table}
\caption[]{\label{tab:numbers} Kaon and pion yields in our
1vs3 samples and estimates of 
background levels. Errors are statistical only.
In each pair of numbers in the table the first number
pertains to kaons and the second to pions. The $\tau$ feed-across
background is included in the $K/\pi$ yields while the hadronic
background 
has already been subtracted.}
\center
\begin{tabular}{llll}
Hadronic & Number of K/$\pi$    & hadronic       & $\tau$ feed- \\
final state  & reconstructed & background, \% & across, \%   \\
\hline
$Kh\pi/\pi\pi\pi$  & $7903\pm 302$/$294780\pm 1184$ 
                           & $3.1/0.5$ & $9.1/11.2$ \\
$Kh\pi\pi^0/\pi\pi\pi\pi^0$ & $719\pm 123$/$55140\pm 680$   
                           & $4.9/0.8$ & $9.8/4.5$ \\
$KK\pi/\pi\pi\pi$ & $2305\pm 211$/$149599\pm 761$  
                           & $5.1/0.4$ & $7.0/14.5$ \\
$KK\pi\pi^0/\pi\pi\pi\pi^0$ & $158\pm 89$/$26915\pm 457$ 
                           & $6.4/0.7$ & $0/7.2$ \\
\end{tabular}
\end{table}

\section{Background}

There are two primary sources of background: 
continuum hadronic events ($e^+e^-\rightarrow q\bar{q} \rightarrow
hadrons$) and non-signal $\tau$ decays (``$\tau$ 
feed-across'').
We estimate hadronic background 
from a continuum hadronic Monte Carlo
sample (using the JETSET v7.3\cite{JETSET} event generator and GEANT
\cite{GEANT} detector simulation code). The kaon and pion momentum spectra 
resulting from q\=q
events that satisfy our selection criteria are found from this 
Monte Carlo sample
and subtracted from the
data $K/\pi$ spectra  prior to fitting for the
$\tau\to KX$ and $\tau\to\pi X$ yields.
The level of hadronic background is shown
in Table \ref{tab:numbers}.

  $\tau$ decay modes containing 
$K^0_S$ mesons are considered feed-across background, because
the major source of 
$\tau$ background to $\tau^-\rightarrow K^-h^+\pi^-(\pi^0)\nu_\tau$ 
is found to be
$\tau^-\rightarrow K^- K^0(\pi^0)\nu_\tau$ decays, in which 
$K^0_S\to\pi^+\pi^-$. There is also contamination of
modes without $\pi^0$'s from modes with $\pi^0$'s, and vice versa, which
we also determine from Monte Carlo simulations, using our measured
branching fractions as inputs.
Three prong decays
with kaons and more than one $\pi^0$ are severely phase
space suppressed and are neglected in this analysis.
The approximate level of $\tau$ background is also given in 
Table \ref{tab:numbers}.

\section{Numerical Results}

We determine the ratio of branching fractions 
relative to the normalizing modes directly from the fitted number of
kaon and pion tracks in the 1vs3 sample.
    For the $\tau^-\rightarrow K^-h^+\pi^-\nu_\tau$ and 
$\tau^-\rightarrow K^-h^+\pi^-\pi^0\nu_\tau$
decay modes, 
each event contributes 2 same-sign tracks to the analyzed
sample of tracks, one of which is a kaon and one a
pion. By
contrast, each $\tau^-\to \pi^-\pi^+\pi^-\nu_\tau$ event
contributes 2 pions. Straightforward algebra can be used
to find a simple expression for the desired ratio of branching
fractions, as outlined in the Appendix.
Using this calculation
we obtain the results for the ratios ($R$) of
branching fractions shown in Table \ref{tab:result}. The first error shown
is statistical and the second is systematic. 
%For each ratio, we 
%multiply by the currently tabulated branching fractions\cite{PDG96} 
%for the normalizing modes to determine an absolute branching fraction for
%each $\tau\to KX$ mode, as indicated in Table \ref{tab:result}.
%
%To determine absolute branching fractions we multiply R by the
%world average branching fractions \cite{PDG96}.

{%\large
\begin{table}
\caption[]{\label{tab:result} Final results for ratios of branching
  fractions and derived absolute branching fractions for all four analyses.}
\center
\begin{tabular}{lll} 
Decay Mode & Ratio definition & Value ($\times 10^{-2}$) \\
\hline
$\tau\rightarrow Kh\pi\nu_\tau$ &
  ${\cal B}(\tau\rightarrow Kh\pi\nu_\tau)
         /{\cal B}(\tau\rightarrow \pi\pi\pi\nu_\tau)$&$5.44\pm 0.21 \pm 0.53$ 
\\
$\tau\rightarrow Kh\pi\pi^0\nu_\tau$ &
 ${\cal B}(\tau\rightarrow Kh\pi\pi^0\nu_\tau) 
         /{\cal B}(\tau\rightarrow \pi\pi\pi\pi^0\nu_\tau)$&$2.61\pm 0.45 \pm 0.42$ 
\\
$\tau\rightarrow KK\pi\nu_\tau$ &
 ${\cal B}(\tau\rightarrow KK\pi\nu_\tau)
         /{\cal B}(\tau\rightarrow \pi\pi\pi\nu_\tau)$&$1.60\pm 0.15 \pm 0.30$ 
\\
$\tau\rightarrow KK\pi\pi^0\nu_\tau$ &
 ${\cal B}(\tau\rightarrow KK\pi\pi^0\nu_\tau)/  
         {\cal B}(\tau\rightarrow \pi\pi\pi\pi^0\nu_\tau)$&$0.79\pm 0.44 \pm 0.16$
\\
%\hline
\end{tabular} 
\end{table}
}

%
%\begin{tabular}{lll} 
%Decay Mode & $R$ ($\times 10^{-2}$) & Absolute ${\cal B}$, \% \\ 
%\hline
%$\tau\rightarrow Kh\pi\nu_\tau$ &
%  $\frac{{\cal B}(\tau\rightarrow Kh\pi\nu_\tau)}
%         {{\cal B}(\tau\rightarrow \pi\pi\pi\nu_\tau)}$=$5.44\pm 0.21 \pm 0.53$ 
%& $0.489\pm 0.019\pm 0.047$\\
%$\tau\rightarrow Kh\pi\pi^0\nu_\tau$ &
% $\frac{{\cal B}(\tau\rightarrow Kh\pi\pi^0\nu_\tau)} 
%         {{\cal B}(\tau\rightarrow \pi\pi\pi\pi^0\nu_\tau)}$=$2.61\pm 0.45 \pm 0.39$ 
%& $0.108\pm 0.019\pm 0.016$\\
%$\tau\rightarrow KK\pi\nu_\tau$ &
% $\frac{{\cal B}(\tau\rightarrow KK\pi\nu_\tau)}  
%         {{\cal B}(\tau\rightarrow \pi\pi\pi\nu_\tau)}$=$1.60\pm 0.15 \pm 0.26$ 
% & $0.144\pm 0.013\pm 0.024$\\ 
%$\tau\rightarrow KK\pi\pi^0\nu_\tau$ &
% $\frac{{\cal B}(\tau\rightarrow KK\pi\pi^0\nu_\tau)}  
%         {{\cal B}(\tau\rightarrow \pi\pi\pi\pi^0\nu_\tau)}$=$0.79\pm 0.44 \pm 0.16$
% & $0.033\pm 0.018\pm 0.007$\\ 
%%\hline
%\end{tabular} 

\section{Systematic errors}

 The breakdown of systematic errors 
%in our calculation of the ratio of branching fractions 
for each decay mode is given in Table \ref{tab:system}. 
\begin{table}[h]
\caption[]{\label{tab:system} Systematic errors}
\center
\begin{tabular}{lllll}
%\hline 
Source  & \multicolumn{4}{c}{Ratio of branching fractions} \\
     &  $\frac{K h\pi\nu}{\pi\pi\pi\nu}$ &  
             $\frac{K h\pi\pi^0\nu}{\pi\pi\pi\pi^0\nu}$ 
               &  $\frac{KK\pi\nu}{\pi\pi\pi\nu}$ &  
                   $\frac{KK\pi\pi^0\nu}{\pi\pi\pi\pi^0\nu}$ \\
\hline 
\hline
Kaon extraction procedure         & 6\%    & 7\% & 9\% & 8\%   \\
MC Model uncertainties               & 2\%    & 6\% & 9\% & 15\%    \\
$\tau$ MC branching fraction uncertainties  &  2\%    &  9\%  & 4\%  & 3\%  \\
MC statistics (for efficiencies)  &  1\%    &  3\%  & 2\%  & 4\%  \\
q\=q background                    & 3\%    &  5\%  & 5\%  & 6\%  \\
Other backgrounds ($2\gamma$, QED,
      beam-gas)                   &  5\%    &  5\%  & 5\%  & 5\% \\
Photon finding/veto                & 4\%    &  6\%  & 10\%  & 6\%  \\
Tracking, Trigger, Tag ID       & cancels  & cancels& cancels  & cancels\\
\hline
Total                             & 10\%    & 16\%& 19\% & 20\%  \\
%\hline
\end{tabular}
\end{table}
The dominant systematic errors arise from the uncertainty in the 
fit procedure used to determine the number of kaons and pions 
in the 1vs3 sample and from the choice of decay models in Monte Carlo
simulation.
The former error is estimated by performing cross-checks on independent
mixtures of kaons and pions with known fractions of
particles of each type. We obtain tagged samples of
kaons and pions using data samples of $\phi \rightarrow K^+ K^-$ or
$D^{*+}\rightarrow D^0 \pi^+$ events, with  the $D^0$ decaying to either
$K^-\pi^+$ or $K^-\pi^+\pi^-\pi^+$.
The $\delta_k$ distributions of kaons and pions are 
added in proportions ranging from 1:1 to 1:25, and the signal extraction
procedure applied. 
The number of fitted kaons is compared to the 
true number of input kaons and a ratio ($N_{fit}\over N_{true}$)
determined. The results of this procedure
are used to estimate the systematic errors 
inherent in the signal extraction procedure.
We then extrapolate the results of this cross-check
to the mixture appropriate to each specific decay mode and assess
the corresponding systematic error of the signal extraction. 
The expected fractions of kaons in the measured decay modes, averaged
over the momentum range $p>$1.5 GeV/$c$,   are 1:22, 1:31, 1:49 and 1:34
for the $\tau\to Kh\pi\nu$,  $Kh\pi\pi^0\nu$,  $KK\pi\nu$ and  
$KK\pi\pi^0\nu$
analyses, respectively.

The MC modeling error listed in Table \ref{tab:system} 
includes the uncertainty in fitting the extracted pion and 
kaon momentum spectra (i.e., extrapolating into the 
$p<$1.5 GeV/$c$ region), as well as the efficiencies for
Monte Carlo events to pass both our event and track selection criteria.
To evaluate this error,
a variety of decay models, both resonant and non-resonant,
are used to determine the possible shapes of the $dN\over dp$ 
spectra and
to recalculate branching ratios.
%\footnote{
%
% Models - footnote - not
%
The following models
were investigated in order to evaluate the systematic error: 
for $\tau^-\to K^-\pi^+\pi^-\nu_\tau$:
$\tau\to K_1(1270)\nu_\tau$, $\tau\to K_1(1400)\nu_\tau$, and phase space;
for $\tau^-\to K^-K^+\pi^-\nu_\tau$:
$\tau\to K^{*0}K\nu_\tau$, $\tau\to\rho(1690)\nu_\tau$ 
($\rho(1690)\to K^{*0}K$), and phase space;
for $\tau^-\to K^-\pi^+\pi^+\pi^0\nu_\tau$:
$\tau\to K\omega\nu_\tau$ and phase space; for
$\tau^-\to K^-K^+\pi^-\pi^0\nu_\tau$:
$\tau\to  K^{*}K^{*0}$, 
$\tau\to\rho(1690)\nu_\tau$ ($\rho(1690)\to K^{*}K^{*0}$), 
and phase space.
To obtain central values, 
% in calculation of efficiencies 
% and extrapolation of momentum spectrum to $P<1.5GeV/c$ region  
the following primary models were
used: the model described in \cite{tauola}
 for the $\tau\to K\pi\pi\nu_\tau$ decay mode
(mostly $K_1(1400)\to K^*\pi$); a mixture of phase space and
$\tau\to K\omega\nu_\tau$ in proportions 75:25 for the
$\tau\to K\pi\pi\pi^0\nu_\tau$ decay mode; the current KORALB \cite{tauola}
 model (including $K^*K$ and $\rho K$) for the
$\tau\to KK\pi\nu_\tau$ decay mode; and a mixture of phase space and
$\tau\to K^*K^{*0}\nu$ in proportions 50:50 for the
$\tau\to KK\pi\pi^0\nu_\tau$ decay mode.
%}
%
%
Because $\tau$ decay modes with kaons are not well
understood theoretically, and because data on such decays is sparse,
this modeling uncertainty is somewhat large. 

To determine the 
systematic error in our feed-across
estimate due to the uncertainty in the
input tau decay branching fractions,
several different samples of generic $\tau$ Monte Carlo were generated
using the KORALB package,
with branching fractions of the components changed within
$\pm 1\sigma$ of the known value. Most feed-across corrections
are determined using branching fractions 
from the Particle Data Group\cite{PDG96}; the magnitude of 
feed-across corrections internal to 
this measurement
(e.g., $\tau^-\to K^-\pi^+\pi^-\pi^0\nu_\tau$ contamination of
$\tau^-\to K^-\pi^+\pi^-\nu_\tau$) are taken from
 the results of this analysis.
The quoted systematic error is derived
from the observed variation of the final results when the input branching
fractions are varied.

We conservatively
assign a 100\% systematic error on the hadronic
background level (see Table \ref{tab:result}), 
since our hadronic simulation
may not accurately model the q\=q background. 
Remaining backgrounds, namely 2-photon events, beam-gas interactions, and
QED background, are assessed by varying the  event and track 
selection requirements
and determined to be less than 5\%.
%To account for systematics related
%to photon-finding, and the neutral energy veto, we assign a 4\%
%systematic error for each mode. 
To account for systematics related
to photon-finding, and the neutral energy veto, we investigate
the dependence of the final results upon the particular values
of the cuts used in $\pi^0$ reconstruction and the photon veto.
This study gives 2-10\% systematic errors (Table \ref{tab:system}).

We assign a MC statistics
 error corresponding to the statistical error on the efficiencies
and feed-across corrections determined from Monte Carlo simulations. 
There are other
systematic effects that cancel in the final ratio of
branching fractions
such as trigger efficiencies, tag identification requirements 
and track-finding systematics.

\section{Summary}

 We have measured the following ratios of branching fractions:
\begin{equation}
    {\cal B}(\tau^-\rightarrow K^-h^+\pi^-\nu_\tau)/
	{\cal B}(\tau^-\rightarrow \pi^-\pi^+\pi^-\nu_\tau)=
            (5.44\pm 0.21 \pm 0.53)\times 10^{-2},
      \label{eq:br1} \\
\end{equation}
\begin{equation}
      {\cal B}(\tau^-\rightarrow K^-h^+\pi^-\pi^0\nu_\tau)/
           {\cal B}(\tau^-\rightarrow \pi^-\pi^+\pi^-\pi^0\nu_\tau)=
             (2.61\pm 0.45 \pm 0.42)\times 10^{-2}, 
      \label{eq:br2}
\end{equation}
\begin{equation}
      {\cal B}(\tau^-\rightarrow K^-K^+\pi^-\nu_\tau)/
         {\cal B}(\tau^-\rightarrow \pi^-\pi^+\pi^-\nu_\tau)=
           (1.60\pm 0.15 \pm 0.30)\times 10^{-2},
      \label{eq:br3}
\end{equation}
\begin{equation}
   {\cal B}(\tau^-\rightarrow K^-K^+\pi^-\pi^0\nu_\tau)/
  {\cal B}(\tau^-\rightarrow \pi^-\pi^+\pi^-\pi^0\nu_\tau)<0.0157(95\% C.L.),
      \label{eq:br4}
\end{equation} 
where the limit is quoted because the value in Table \ref{tab:result}
is not statistically significant. Contributions to both denominator
and numerator from $\tau\to K^0_SX$;  $K^0_S(\to\pi^+\pi^-)$ have 
been excluded. If we instead
normalize to $\tau\rightarrow h^-h^+h^-(\pi^0)\nu_\tau$, the
corresponding ratio of branching fractions are:
\begin{equation}
    {\cal B}(\tau^-\rightarrow K^-h^+\pi^-\nu_\tau)/
         {\cal B}(\tau^-\rightarrow h^-h^+h^-\nu_\tau)=
(5.16\pm 0.20 \pm 0.50)\times 10^{-2},
      \label{eq:br5}
\end{equation}
\begin{equation}
 {\cal B}(\tau^-\rightarrow K^-h^+\pi^-\pi^0\nu_\tau)/
        {\cal B}(\tau^-\rightarrow h^-h^+h^-\pi^0\nu_\tau)=
(2.54\pm 0.44 \pm 0.39)\times 10^{-2}, 
      \label{eq:br6}
\end{equation}
\begin{equation}
  {\cal B}(\tau^-\rightarrow K^-K^+\pi^-\nu_\tau)/
      {\cal B}(\tau^-\rightarrow h^-h^+h^-\nu_\tau)=
(1.52\pm 0.14 \pm 0.29)\times 10^{-2}, and
      \label{eq:br7}
\end{equation}
\begin{equation}
  {\cal B}(\tau^-\rightarrow K^-K^+\pi^-\pi^0\nu_\tau)/
    {\cal B}(\tau^-\rightarrow h^-h^+h^-\pi^0\nu_\tau)<0.0154.
      \label{eq:br8}
\end{equation}
    Subtracting (\ref{eq:br7}) from (\ref{eq:br5}) and the central
value of (\ref{eq:br8}) from (\ref{eq:br6}) we find:
\begin{equation}
  {\cal B}(\tau^-\rightarrow K^-\pi^+\pi^-\nu_\tau)/
         {\cal B}(\tau^-\rightarrow h^-h^+h^-\nu_\tau)=
(3.64\pm 0.24 \pm 0.58)\times 10^{-2},
      \label{eq:br9}
\end{equation}
and
\begin{equation}
  {\cal B}(\tau^-\rightarrow K^-\pi^+\pi^-\pi^0\nu_\tau)/
        {\cal B}(\tau^-\rightarrow h^-h^+h^-\pi^0\nu_\tau)=
  (1.77\pm 0.62\pm 0.42)\times 10^{-2}. 
      \label{eq:br10}
\end{equation}
Using the CLEO measurements of the decay channels
$\tau^-\to h^-h^+h^-\nu_\tau$ and 
$\tau^-\to h^-h^+h^-\pi^0\nu_\tau$ \cite{cleo3h}
 for the denominator, and
% taken from \cite{cleo3h}, and 
Eqns. (\ref{eq:br7})-(\ref{eq:br10}), we find the branching fractions
given in Table \ref{tab:isospin-Kpp}.

\section{Discussion}

The $\tau^-\to K^-\pi^+\pi^-\nu_{\tau}$
decay mode
is believed to occur predominantly through coupling to the
axial-vector mesons $K_1(1270)$ and $K_1(1400)$. 
The numerical prediction
for the branching fraction of this decay mode calculated by Finkemeier and
Mirkes\cite{mirkes} is $0.77\%$, 
more than twice as large as both our result as well as the result
of ALEPH given in Table \ref{tab:isospin-Kpp}.
Another theoretical prediction, 0.18\% by Li\cite{li}, is consistent
with present measurements.

\begin{table}
\caption[]{\label{tab:isospin-Kpp}
  Recent measurements of $\tau\to Kh\pi(\pi^0)\nu_\tau$  decay modes.}
\center
\begin{tabular}{lll}
$\tau$ decay mode & Measurement & Branching fraction, $10^{-2}$ \\
\hline
$\tau^-\to\bar{K^0}\pi^-\pi^0\nu_\tau$ &
             ALEPH\cite{aleph-97167}   &$0.294\pm 0.073\pm 0.037$ \\
          &  CLEO\cite{kshort_paper}   &$0.417\pm 0.058\pm 0.044$\\
 & & \\
$\tau^-\to K^-\pi^+\pi^-\nu_{\tau}$    &
             ALEPH\cite{aleph_pre}     &$0.214\pm 0.037\pm 0.029$ \\
          &  This analysis             &$0.346\pm 0.023 \pm 0.056$\\
	  & Theory                     & 0.77 in \cite{mirkes},
                                             0.18 in \cite{li} \\
 & & \\
$\tau^-\to K^-\pi^0\pi^0\nu_\tau$      &
             ALEPH\cite{aleph-k0pp0}   &$0.08\pm 0.02\pm 0.02$\\
          &  CLEO\cite{cleo-kp0p0}     &$0.14\pm 0.10\pm 0.03$\\
\hline 
$\tau^-\to K^-K^0\pi^0\nu_\tau$       &
       ALEPH\cite{aleph-97167}        &$0.152\pm 0.076\pm 0.021$\\
   &   CLEO\cite{kshort_paper}        &$0.145\pm 0.036\pm 0.020$ \\ 
 & & \\
$\tau^-\to K^-K^+\pi^-\nu_{\tau}$     &
       ALEPH\cite{aleph_pre}          &$0.163\pm 0.021\pm 0.017$ \\
  &    This analysis                  &$0.145\pm 0.013 \pm 0.028$ \\
  &    Theory \cite{mirkes}           & 0.22 \\
 & & \\
$\tau\to  K^0_S K^0_L\pi^-\nu_\tau$  &
        ALEPH\cite{aleph-97167}        &$0.101\pm 0.023\pm 0.013$ \\ 
$\tau\to  K^0_S K^0_S\pi^-\nu_\tau$  &
        CLEO\cite{kshort_paper}     &$0.023\pm 0.005\pm 0.003$ \\
     & ALEPH\cite{aleph-97167}      &$0.026\pm 0.010\pm 0.005$ \\
$\tau\to  K^0\bar{K^0}\pi^-\nu_\tau$  &
       L3\cite{l3}                   &$0.31\pm 0.12\pm 0.04$ \\
\hline
$\tau^-\to K^-\pi^+\pi^-\pi^0\nu_{\tau}$    &
             ALEPH\cite{aleph_pre}     &$0.061\pm 0.039\pm 0.018$ \\
          &  This analysis             &$0.075\pm 0.026 \pm 0.018$\\
\hline
$\tau^-\to K^-K^+\pi^-\pi^0\nu_{\tau}$    &
             ALEPH\cite{aleph_pre}     &$0.075\pm 0.029\pm 0.015$ \\
          &  This analysis             &$0.033\pm 0.018 \pm 0.007$\\
\end{tabular}
\end{table}

In contrast
to the $K\pi\pi$ mode, tau decays involving the $KK\pi$ final state
may occur      through either the vector or axial vector currents. 
Theoretical predictions for the relative amounts of V and A vary 
considerably  \cite{gomez,braaten,decker,li}.
One can use isospin
symmetry to relate the $K^-K^0\pi^0$, $K^-K^+\pi^-$ and $K^0\bar{K^0}\pi^-$
tau decay modes. The ratio of the branching fractions of these decay modes
should be 2:1:1 if $\tau^-\to K^-K^+\pi^-\nu_{\tau}$ proceeds exclusively
through the $\rho\pi$ intermediate state 
or 1:1:1 if this decay proceeds through $K^*K$. 
The experimental results for these decay modes are given in 
Table \ref{tab:isospin-Kpp}. The decay rate of 
$\tau\to  K^0\bar{K^0}\pi^-\nu_\tau$ can be inferred from the
 ALEPH's measurement for  ${\cal B}(\tau\to K^0_S K^0_L\pi^-\nu_\tau)$
and the combined measurement of CLEO and ALEPH  
of ${\cal B}(\tau\to K^0_S K^0_S\pi^-\nu_\tau)$
(Table \ref{tab:isospin-Kpp}):
\begin{eqnarray} \nonumber
   {\cal B}(\tau\to  K^0\bar{K^0}\pi^-\nu_\tau) = &  
         {\cal B}(\tau\to K^0_S K^0_L\pi^-\nu_\tau) + 
         2{\cal B} (\tau\to K^0_S K^0_S\pi^-\nu_\tau)  =\\ \nonumber 
  & =     (0.149\pm 0.024 \pm 0.014)\times 10^{-2}          
\end{eqnarray}
% These results
%imply $(0.202\pm 0.046\pm 0.026)\times 10^{-2}$ and 
%$(0.096\pm 0.016\pm 0.012)\times 10^{-2}$
%for $\tau\to  K^0\bar{K^0}\pi^-\nu_\tau$,
%provided that the 
%$K^0\bar{K^0}$ system evolves incoherently.
%
%
Comparison of these numbers with
the isospin-predicted ratios indicates that the bulk of
$KK\pi$ production occurs through the vector $K^*K$ intermediate state.
This conclusion is consistent with the direct measurement of
${\cal B}(\tau\to K^*K\nu_\tau)\over
{\cal B}(\tau\to K^-K^+\pi^-\nu_\tau)$=0.87$\pm$0.13 by ALEPH\cite{aleph_pre}.

We can also interpret the available measurements for $\tau\to KK\pi\nu$ 
to determine the relative couplings of the $\tau$ to the strange vector
or strange axial vector currents by taking advantage of isospin
relations, as in \cite{rouge}. If we calculate the ratios
\begin{equation}
 R_v = 1/R_a = \frac{{\cal B}_{K^-\bar{K^0}\pi^0}}
              {2{\cal B}_{K^-K^+\pi^-} - {\cal B}_{K^-\bar{K^0}\pi^0}},~~~~~ 
  R =  \frac{2{\cal B}_{K_S^0K_S^0\pi^-}}{{\cal B}_{K_S^0K_L^0\pi^-}} 
\end{equation}
{\flushleft{
then $R\approx R_v$ indicates vector dominance while
$R\approx R_a$ implies axial vector dominance. Combining 
the available measurements
from Table \ref{tab:isospin-Kpp} we obtain $R = 0.48^{+ 0.20}_{-0.14}$, 
$R_v = 0.90^{+0.69}_{-0.37}$ and $R_a = 1.11^{+0.75}_{-0.50}$.}}
The asymmetric errors are defined
so that the probability to obtain a measurement within one
standard deviation is equal to $68\%$.
%
%
%In both methods, comparison of the results with
%the isospin-predicted ratios indicates that the bulk of
%$KK\pi$ production occurs through the vector $K^*K$ intermediate state.
%This conclusion is consistent with the direct measurement of
%${\cal B}(\tau\to K^*K\nu_\tau)\over
%{\cal B}(\tau\to K^-K^+\pi^-\nu_\tau)$=0.87$\pm$0.13 by ALEPH\cite{aleph_pre}.
Although the values of $R$, $R_a$ and $R_v$ favor the 
vector $K^*K$ state, the results of this method remain inconclusive
due to the large errors and the proximity of $R_V$ and $R_a$ to 1. 
In addition,
the value of $R$, while being closer to $R_v$,
 is lower than both $R_a$ and $R_v$,
in contrast to the expectation that $R$ should assume a value between
$R_a$ and $R_v$. If this situation does not resolve itself as
errors are reduced, some of the assumptions in the derivation
of these relations may have to be re-examined.
%%This may indicate, for example, that
%% the isospin relations do not hold  exactly for 
%%$\tau\to K K\pi\nu_\tau$ and/or
%% that incoherence of the $K^0\overline{K}^0$
%% system cannot be assumed.
%Note that the closeness of $R_v$ and $R_a$
%to one reduces the sensitivity of this method to the ratio of
%$V/(V+A)$. 
More precise measurements of the 
different $KK$ isospin combinations in
$\tau\to K K\pi\nu_\tau$ should offer some clarification.

%    The measurements presented herein complement other information on the
%$KK\pi$ final state and provide information on the isospin structure of this
%channel.  
%The values of 
%${\cal B}(\tau^-\to K^0\overline{K}^0\pi^-\nu_\tau)$ 
%calculated from
%${\cal B}(\tau^-\to K^0_S K^0_L\pi^-\nu_\tau)$ and 
%${\cal B}(\tau^-\to K^0_S K^0_S\pi^-\nu_\tau)$ 
%differ by approximately two standard
% deviations. 
%In addition, the value of $R$, while being closer to $R_v$,
% is lower than both $R_a$ and $R_v$,
%in contrast to the expectation that $R$ should assume a value between
%$R_a$ and $R_v$. This may indicate, for example, that
% the isospin relations do not hold  exactly for 
%$\tau\to K K\pi\nu_\tau$ and/or
% that incoherence of the $K^0\overline{K}^0$
% system cannot be assumed.
%Note that the closeness of $R_v$ and $R_a$
%to one reduces the sensitivity of this method to the ratio of
%$V/(V+A)$. More precise measurements of the 
%different $KK$ isospin combinations in
%$\tau\to K K\pi\nu_\tau$ should offer some clarification.

   The theoretical prediction for ${\cal B}(\tau^-\to K^-K^+\pi^-\nu_{\tau})$
%decay mode
 is $\sim0.2\%$ in \cite{mirkes}.
Our measurement is consistent with  this value 
and the recent ALEPH measurement of this mode \cite{aleph_pre}.  

The theory for the  $\tau$ decays 
$\tau^-\to K^-\pi^+\pi^-\pi^0\nu_{\tau}$ and 
 $\tau^-\to K^-K^+\pi^-\pi^0\nu_{\tau}$
is more difficult to formulate
than that for the 3-meson decays discussed above due to the
substantially larger number of possible intermediate states. 
Li\cite{li} has calculated $\tau\to\omega K\nu_\tau$=0.025\%, 
which, if correct, would account for approximately 1/3 of our total observed
rate for the $\tau^-\to K^-\pi^+\pi^-\pi^0\nu_{\tau}$.
Explicit
measurements of the substructure in $\tau\to\pi^-\pi^+\pi^-\pi^0\nu_\tau$,
coupled with these results may help to resolve the nature of these
four-meson decays.

% (Table \ref{tab:result}) and 
%ALEPH (${\cal B}(\tau^-\to K^-\pi^+\pi^-\pi^0\nu_{\tau})=
%(0.061\pm0.039\pm0.018)\times 10^{-2}$ and
%${\cal B}(\tau^-\to K^-K^+\pi^-\pi^0\nu_{\tau})=
%(0.075\pm0.029\pm0.015)\times 10^{-2}$) may provide guidance for future
%progress in this sector.

%------------- 

\vspace{1cm}

\centerline{\bf ACKNOWLEDGMENTS}
\smallskip
We gratefully acknowledge the effort of the CESR staff in providing us with
excellent luminosity and running conditions.
J.R. Patterson and I.P.J. Shipsey thank the NYI program of the NSF, 
M. Selen thanks the PFF program of the NSF, 
M. Selen and H. Yamamoto thank the OJI program of DOE, 
J.R. Patterson, K. Honscheid, M. Selen and V. Sharma 
thank the A.P. Sloan Foundation, 
M. Selen and V. Sharma thank Research Corporation, 
S. von Dombrowski thanks the Swiss National Science Foundation, 
and H. Schwarthoff thanks the Alexander von Humboldt Stiftung for support.  
This work was supported by the National Science Foundation, the
U.S. Department of Energy, and the Natural Sciences and Engineering Research 
Council of Canada.

%----------------- Bibliography

\newpage
\section{Appendix: Derivation of Relative Ratio of Branching Fractions}
%
%  Expression for the ratio of branching fractions
%
The expression
for the ratio of branching fractions of
$\tau\rightarrow Kh\pi(\pi^0)\nu_\tau$ and 
    $\tau\rightarrow \pi\pi\pi(\pi^0)\nu_\tau$ decays
 is straightforward to derive.
Each $\tau\rightarrow Kh\pi(\pi^0)\nu_\tau)$ event
contributes one kaon and one pion into the analyzed sample of
tracks and each $\tau\rightarrow \pi\pi\pi(\pi^0)\nu_\tau)$
contributes 2 pions. A system of linear equations can be written
from which follows the formula:
\[ \frac{{\cal B}(\tau\rightarrow Kh\pi(\pi^0)\nu_\tau)}
         {{\cal B}(\tau\rightarrow \pi\pi\pi(\pi^0)\nu_\tau)} = 
  \frac{N_{Kh\pi(\pi^0)}}{N_{\pi\pi\pi(\pi^0)}}=
          \frac{N_K^{fit}\phi_{Kh\pi}}
           {\epsilon_K \phi_{\pi\pi\pi} N_{\pi}^{fit}-
               \epsilon_{\pi fk}\phi_{Kh\pi}N_{K}^{fit}}
  \frac{2\epsilon_{\pi}\epsilon_{\pi\pi\pi}}
       {\epsilon_{Kh\pi}}
\]
where $N_K^{fit}$ and $N_{\pi}^{fit}$ are the fitted numbers of kaons
and pions for a given 1vs3 sample, $\epsilon_K$, 
$\epsilon_{\pi fk}$ and  $\epsilon_{\pi}$ are the efficiencies
for a hadron (kaon or pion from $Kh\pi(\pi^0)$ decay or pion from 
$\pi\pi\pi(\pi^0)$ decay) to pass our track selection requirements,
%(on $\cos{\theta}$ and $N_{hit}$), 
$\epsilon_{Kh\pi}$ and 
$\epsilon_{\pi\pi\pi}$ are the efficiencies for the indicated events 
to pass our
1vs3 event selection cuts, and $\phi_{Kh\pi}$ and $\phi_{\pi\pi\pi}$
represent the $\tau$ feed-across corrections. 

    The expression for the decay modes $\tau^-\rightarrow K^-K^+\pi^-\nu_\tau$
and $\tau^-\rightarrow K^-K^+\pi^-\pi^0\nu_\tau$ is simpler
since only one track is taken from each event:
\[ \frac{{\cal B}(\tau\rightarrow KK\pi(\pi^0)\nu_\tau)}
         {{\cal B}(\tau\rightarrow \pi\pi\pi(\pi^0)\nu_\tau)} = 
  \frac{N_{Kh\pi(\pi^0)}}{N_{\pi\pi\pi(\pi^0)}}=
          \frac{N_K^{obser}\cdot \phi_{Kh\pi} \epsilon_\pi 
                              \epsilon_{\pi\pi\pi}} 
               {N_\pi^{obser}\cdot \phi_{\pi\pi\pi} \epsilon_K 
                              \epsilon_{Kh\pi}} 
\]
where the efficiencies and feed-across corrections have the same
meaning as in the previous formula. 

The values of track efficiencies
 $\epsilon_K$,  $\epsilon_{\pi fk}$ and  $\epsilon_{\pi}$
are $\sim 90\%$ 
and the event efficiencies  $\epsilon_{Kh\pi}$,
$\epsilon_{\pi\pi\pi}$ are $\sim 36\%$ for decays without
$\pi^0$'s and $\sim 14\%$ for decays with $\pi^0$.
%
%  End of text
%

\end{document}